# On the Value of a Social Network

Sandeep Chalasani

**Abstract**

Different laws have been proposed for the value of a social network. According to Metcalfe's law, the value of a network is proportional to $n^2$ where $n$ is number of users of the network, whereas Odlyzko *et al* propose on heuristic grounds that the value is proportional to *n log n*, which is the Zipf's law. In this paper we have examined scale free, small world and random social networks to determine their value. We have found that the Zipf's law describes the value for scale free and small world networks although for small world networks the proportionality constant is a function of the probability of rewiring. We have estimated the function associated with different values of rewiring to be described well by a quadratic equation. We have also shown experimentally that the value of random networks lies between Zipf's law and Metcalfe's law.

**Introduction**

Social networks are structures consisting of individuals or organizations that create powerful ways of communicating and sharing information. Millions of people use social networking websites like MySpace, Facebook, Bebo, Orkut and Hi5. The question of their value related to size is an important problem in computer science [1, 2], both from the point of view of connectivity and that of business investment.

Social networks connect people and the cost involved in connecting is low, which benefits businesses and institutions. These networks are important in customer relationship management, and they serve as online meeting places for professionals. Virtual communities allow individuals to be easily accessible. People establish their real identity in a verifiable place, these individuals then interact with each other or within groups that share common business interests and goals.

False and exaggerated estimates of the value of a social network can have significant implications for technology investors. Until the IT bubble burst in 2001, it had been common to estimate the market value of a social network based on Metcalfe's law which says that a value of a network is proportional to the square of the size of network [1]. Recently Odlyzko and his collaborators have argued against Metcalfe's law saying that it is significantly overestimates value and they have suggested that the value of a general communication network of size *n* grows according to *n log n*, which is Zipf's law [2, 27].

An important claim has been made by anthropologist Robin Dunbar [3, 4] on the extent of connectivity in effective social organizations. He argued that the size of the brain is correlated with the complexity of function and developed an equation, which works for most primates, that relates the neocortex ratio of a particular species - the size of the neocortex relative to the size of the brain – to the largest size of the social group.  For humans, the max group size is 147.8, or about 150. This represents Dunbar's estimate of the maximum number of people who can be part of a close social relationship [4].

Support for Dunbar's ideas come from the community of Hutterites, followers of the sixteenth century Jakob Hutter of Austria, who are pacifists and believe in community property and live in a shared community called colony. Several thousand Hutterites relocated to North America in the late 19th century and their colonies are mostly rural [3,4].  A colony consists of about 10 to 20 families, with a population of around 60 to 150. When the colony's population approaches the upper figure, a *daughter* colony is established.

Dunbar's ideas can be taken to be an indication of the idea that most social networks are "small world" networks [3, 4, 5, and 9]. Small world networks exhibit clustering and small characteristic path lengths that seem to capture many features of social computing networks.  We are interested in relating value to size in such networks.

In this paper we propose to investigate the value of a social network with respect to the probability mechanism underlying its structure. Specifically we compute the value for small world networks and scale free networks. We provide evidence in support of the value to be given by Zipf's law.

**Zipf's Law**

Zipf's law is an empirical law originally proposed for words in a large text and it states that given some corpus of natural language utterances, the frequency of any word is inversely proportional to its rank in the frequency table. The most frequent word will occur approximately twice as often as the second most frequent word, which occurs twice as often as the fourth most frequent word etc. In the network context, if the value of the most important member to user *A* is taken to be proportional to 1; that of the second most important member is proportional to ½, and so on. For a network that has n members, this value to the user *A* will be proportional to *1 + 1/2 + 1/3 +…+ 1 / (n-1)*, which approximates to *log n*. Given that the number of users is *n*, the total value of the network is proportional to *n log n*.

Metcalfe's law took the value of the network to be proportional to its connectivity, since the total number of connections in a network of *n* users is *n (n-1)* or about $n^2$. In practice many users will be connected socially only to a fraction of all the users though the networks provide a full connectivity of $n^2$. Reed's law [7] is based on the insight that in a communication network as flexible as internet, in addition to linking pairs of members. With *n* participants, there are $2^n$ possible groups, and if they are all equally valuable, the value of the network grows like $2^n$.

**Probabilistic Random Networks**

    We consider probabilistically generated social networks. These networks are based on the variable binomial distribution in which sets of nodes are connected to other nodes with different probability distributions. A sample random network with 12 nodes and 62 connections is shown

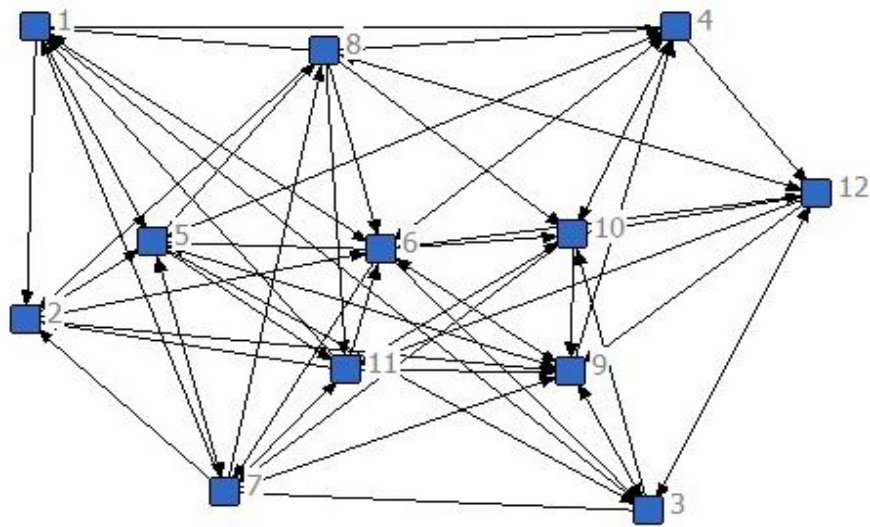

in Fig. 1.

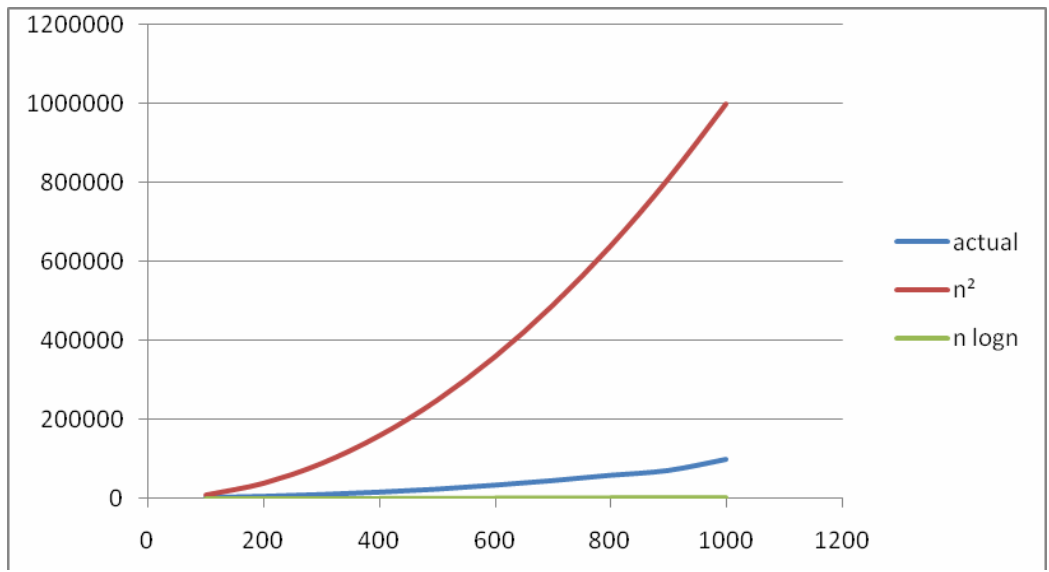

Fig.2 Graph showing the values of example networks compared with $n^2$ and $n \log n$

    Figure 2 shows the values of the network in comparsion with values $n^2$ and $n \log n$. The number of nodes in the network is the X-axis and the associated value for each node is in the Y-axis. From this observation we clearly understand that the actual value of the network lies somewhere in between the values $n^2$ and $n \log n$.

**Small World Networks**

We have simulated small world networks and and the value associated to coressponding graphs are observed on an average case. We generated a Watts-Strogatz small world network consisting of *N* nodes [17, 18, 21]. Each node is directly connected to *k* immediate neighbours that are located symmetrically in the ring lattice on two sides of the node.

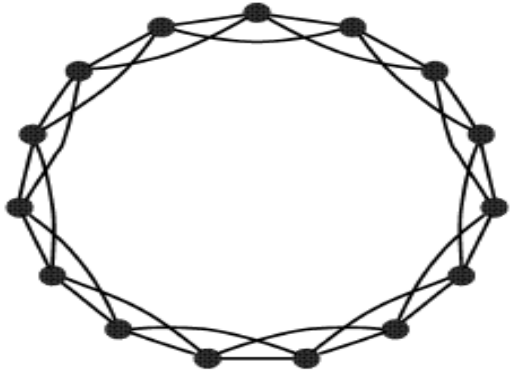

Fig. 3 Watts Strogatz ring lattice for a small world networks with 15 Nodes and 4 local contacts for each node

A small world network is generated by "rewiring" the basic network i.e. ring lattice. Rewiring at each node consists of redirecting one of the outgoing arcs at the node to some other destination node. The extent of rewiring is controlled by probability *p*. We generate a random number which is uniformly distributed and check whether the generated random number is less than or greater than the given probability. If the random number generated is less than the assumed probability we rewire an arc, otherwise the arc is left unchanged.

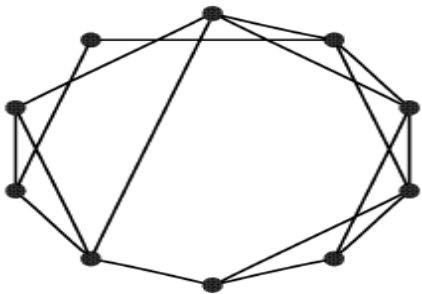

Fig.4 Small world network with the probability of rewiring is *p*=0.08

As we increase the value of the *p* from 0 to 1.0 we see a randomly rewired graph almost all the nodes connected differently

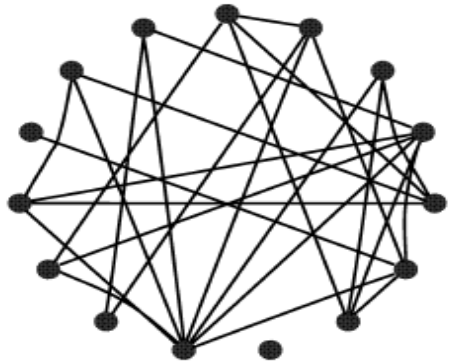

Fig. 5 Small world network with probability of rewiring *p*=1.0.

**Value of Small World Networks**

Several small world networks with different number of nodes are generated using different binomial distribution for random number generation and with variable probability values for the rewiring. In every network for every node we count the number of other nodes to which it is connected and the total value of the network is estimated. We considered several repetitions of the generations and the average case value is considered.

| No. of Nodes | Calculated Value | Metcalfe's | Odlyzko |
|---|---|---|---|
| 100 | 747 | 10000 | 200 |
| 90 | 689 | 8100 | 176 |
| 80 | 634 | 6400 | 152 |
| 70 | 518 | 4900 | 129 |
| 60 | 490 | 3600 | 107 |
| 50 | 345 | 2500 | 86 |
| 40 | 256 | 1600 | 64 |

**Table. 1.** Comparison between the calculated value, $n^2$ and $n \log n$ values of small world networks with different sizes with *p*=0.18

The graph showing the different value curves as observed for a small world network with a probability of rewiring *p* as 0.18 and 0.32 in Figures 6 and 7.

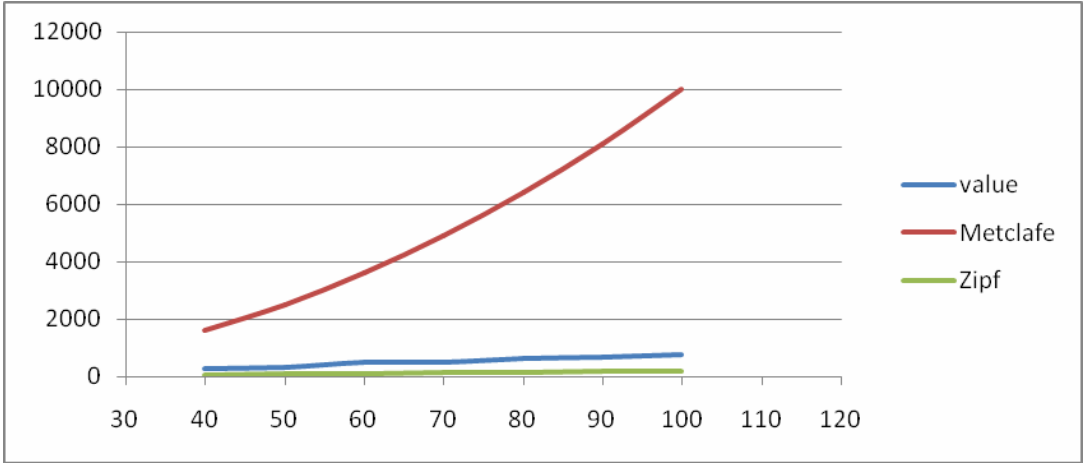

Fig. 6 Graph comparing the Values of small world network with a *p*=0.18

| No. of Nodes | Calculated value | Metcalfe | Zipf's |
|---|---|---|---|
| 100 | 1296 | 10000 | 200 |
| 90 | 1221 | 8100 | 176 |
| 80 | 1025 | 6400 | 152 |
| 70 | 830 | 4900 | 129 |
| 60 | 730 | 3600 | 107 |
| 50 | 522 | 2500 | 85 |
| 40 | 384 | 1600 | 64 |

**Table.2.** Comparison between the calculated Value, $n^2$ and $n \log n$ values of small world networks with different sizes with probability of rewiring *p*=0.32

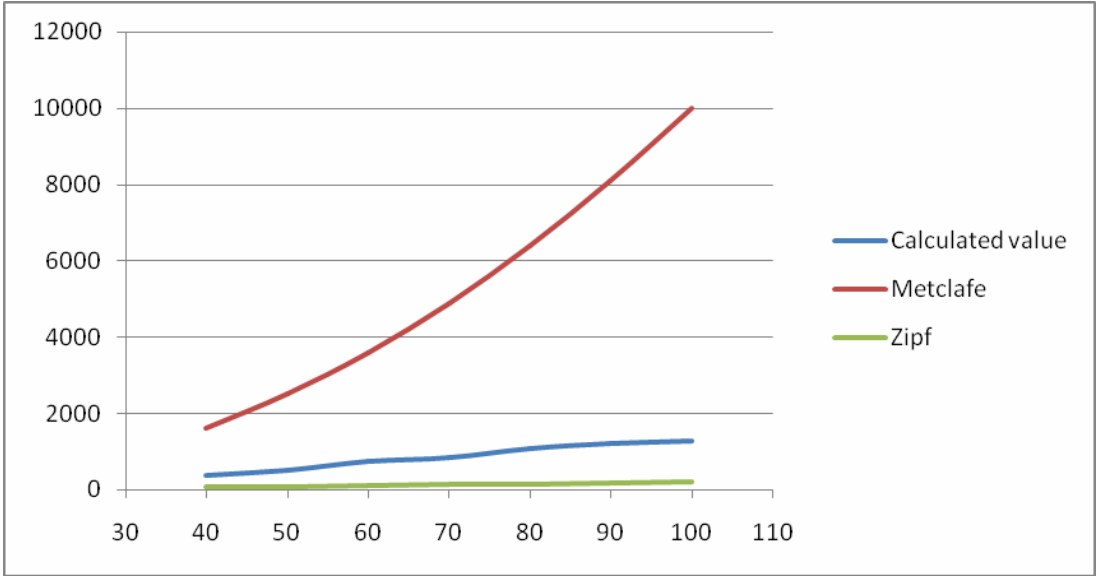

Fig. 7 Graph comparing the values of small world network with a *p*=0.32

We observe that in Table1 and Table 2 that the calculated value is 4 and 6 times that of *n log n* respectively. We established a relation between probability and number of times the calculated value is more than that of *n log n* as and where the functional relationship is given by the following quadratic relationship

$$Y=12.045x^2 + 6.59x+2.5533$$

The regression value for this quadratic function is quite close to 1.

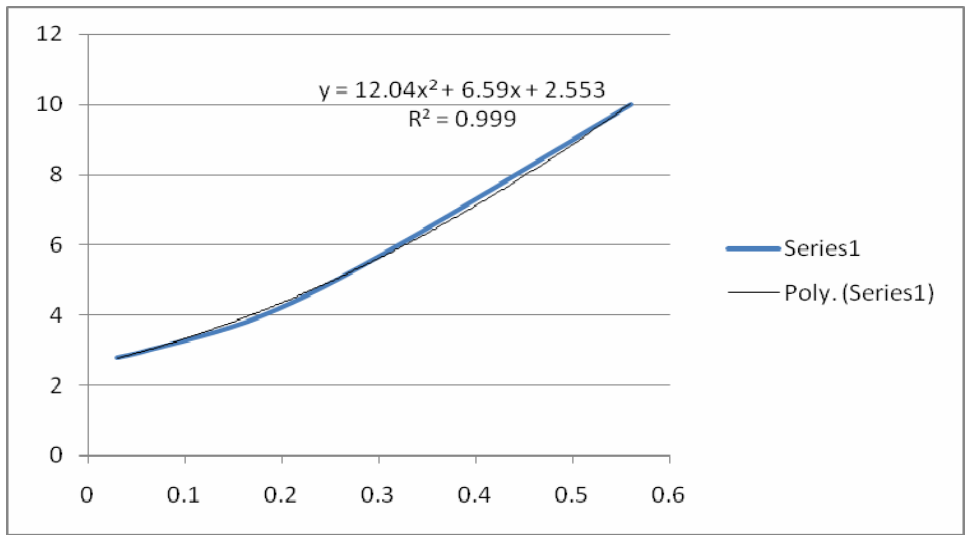

Fig. 8 Graph showing the relation between probability and the no. of times calculated value is more than *n log n*.

**Scale Free Networks**

A scale free network is network whose degree distribution follows a power law. We have simulated Barabasi and Albert (B-A) [10, 11] model of scale free networks. We generated a network of small size, and then used that network as a seed to build a greater sized network, continuing this process until the actual desired network size is reached. The initial seed used need not have scale free properties, while the later seeds may happen to have these properties.

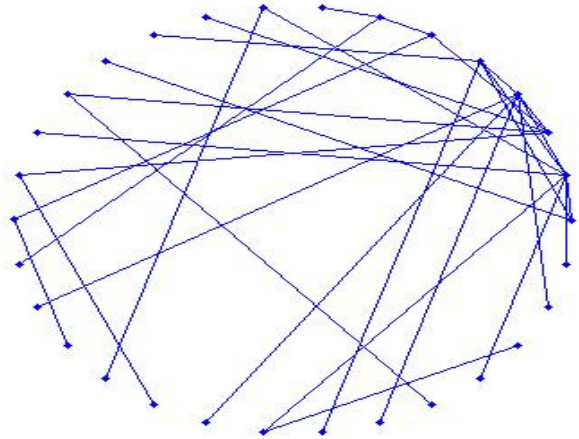

Fig. 9 B-A Scale Free graph with 30 nodes

We can draw a best fit line to the frequency of degrees distribution of the nodes. Degree is the number of links that connect to and fro a single node. For scale free networks, the frequency of degrees distribution forms a power- law curve, with a exponent usually between -2 and -3.

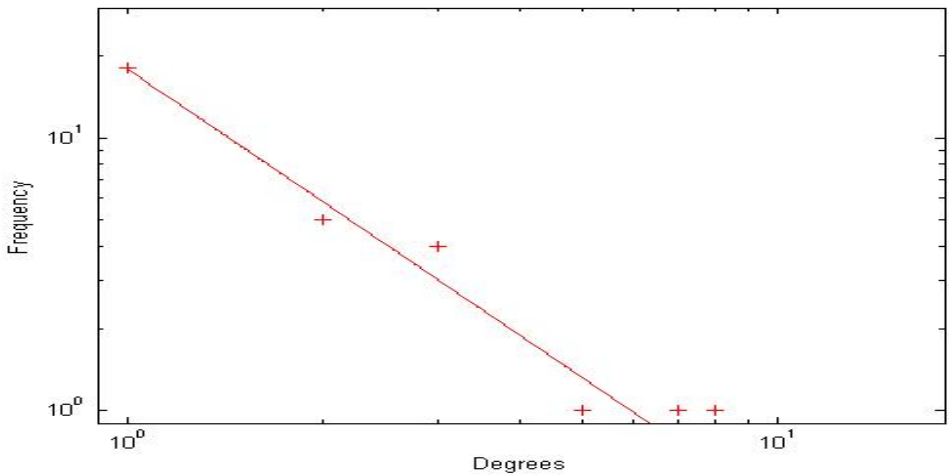

Fig.10 Power-law curve for the small world network in Fig.9.

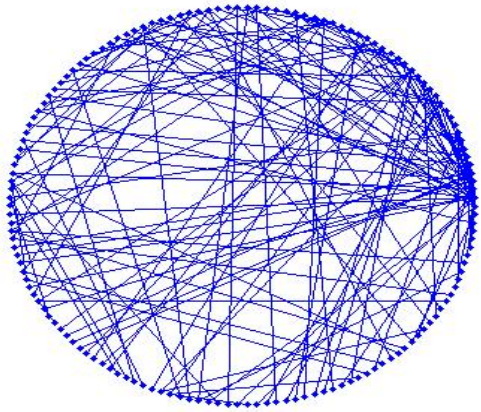

Fig. 11 B-A small world network with 150 nodes

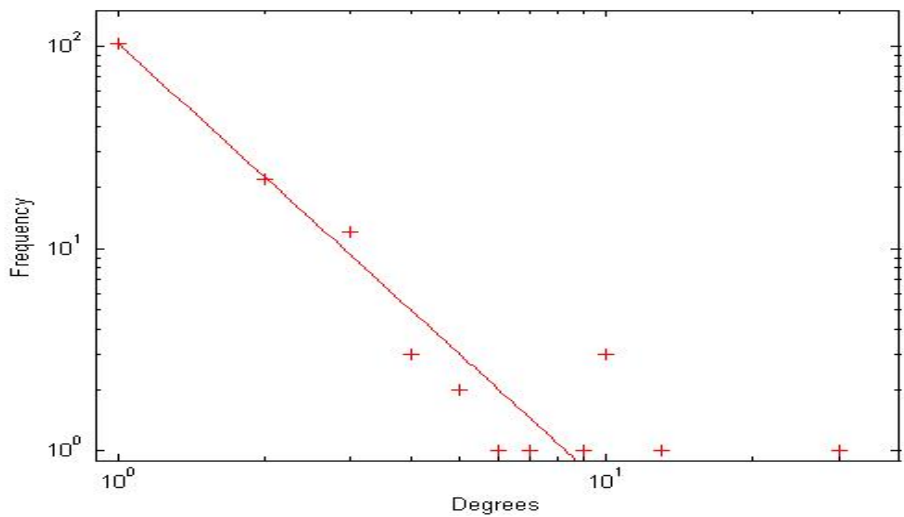

Fig. 12 Graph showing the power law distribution for the small world network in Fig.11

**Value of Scale Free Networks**

Several of these scale free networks are generated and the average case for the value calculation is taken into account. These scale free networks follow the Power law and therefore the values associated with them correspond to n log n.

| No. of nodes | Calculated value | Odlyzko n log n |
|---|---|---|
| 30 | 60 | 44.31 |
| 40 | 80 | 64.082 |
| 50 | 100 | 84.94 |
| 60 | 120 | 106.689 |
| 70 | 140 | 129.156 |
| 80 | 160 | 152.247 |
| 90 | 180 | 175.88 |
| 100 | 200 | 200 |

Table 3. Showing the Values Scale Free networks with different nodes

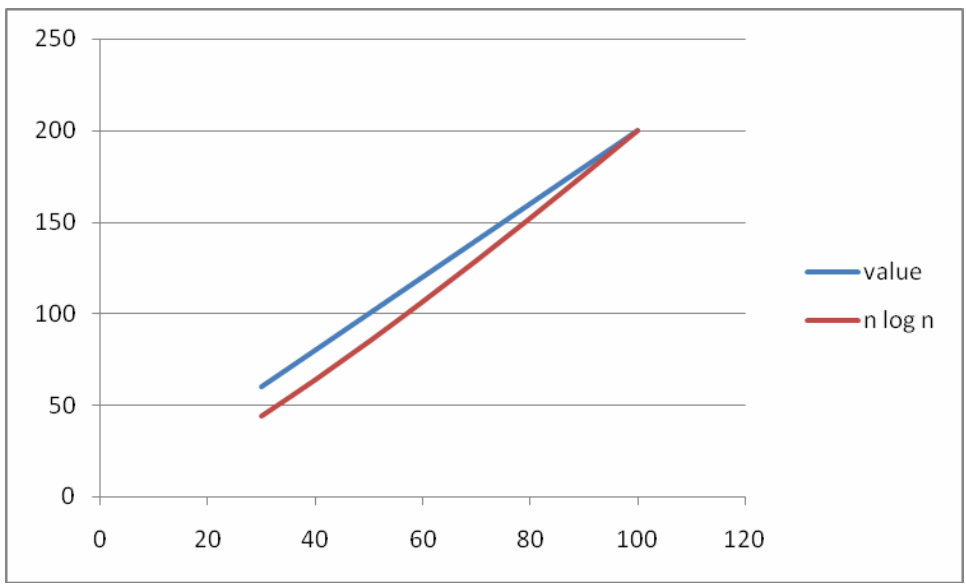

Fig. 13 Graph comparing the value of scale free network and n log n

This is also seen in Figure 13. We conclude that the property of being scale free captures the underlying foundation of the Zipf's law.

# Conclusion

We have demonstrated that the Zipf's law, originally proposed on heuristic grounds, is valid for scale free and small world networks. We have shown empirically that the expression of value for a Watts- Strogatz small world network of *n* nodes is

$$f(p) \, n \log n$$

$$f(p) = 12.054p^2 + 6.59p + 2.5533$$

where *p* is the probability of rewiring. We have computed the value of *f (p)* for various *p* and found that the quadratic function provides an excellent fit. We believe that this is the first study broadly validating the heuristic claim of Odlyzko *et al* on the value of social networks.

Although no specific relationship between size and value can be fixed for random networks, our simulation shows that this value lies between Zipf's law and Metcalfe's law.

As future study one would like to determine if non-Watts-Strogatz small world networks also follow the Zipf's law.